# Ultra-High Insensitive Terahertz Absorber for TE and TM Polarizations Based on Graphene Disks Configuration

Omid Mohsen Daraei[1], Milad Rastkar Mirzaei[1,2], Kiyanoush Goudarzi[1,2], Mohammad Bemani[1*]

[1] Department of Electrical and Computer Engineering, University of Tabriz, Tabriz, Iran

[2] Quantum Photonic Research Lab, University of Tabriz, Tabriz, Iran

*Corresponding author: *bemani@tabrizu.ac.ir*

## Abstract

In this paper, we have proposed ultra-broadband terahertz absorber with three layers of periodic graphene disks arrays (PGDAs). A gold layer was utilized as backside reflector and TOPAS as dielectric spacers. Circuit model of PGDAs and the developed Transmission Line Model (TLM) for the absorber is utilized to design the structure using impedance matching concept. The validity of the results of the device was confirmed by Finite Element Method (FEM). The structure absorbs both TE and TM polarizations. Furthermore, the obtained normalized absorption bandwidth is 148% with a bandwidth of 2.64 THz in 1.78 THz central frequency which is to best of our knowledge the best report in two-dimensional graphene-based structures.

## 1. Introduction

In recent years, the terahertz band has been subject to intense research due to its numerous applications in imaging, communication, security, and spectroscopy [1]–[4]. Metamaterial terahertz absorbers have wide applications in terahertz and infrared imaging, bio-sensing, photodetection, wave manipulation, energy harvesting, radiative cooling, and chemical detection, owing to their remarkable advantages like near unity absorption [5]–[7]. Nevertheless, most of the reported absorbers suffer from fixed working frequency and narrow-bandwidth absorption [5]. To resolve these drawbacks, graphene-based terahertz absorbers have been proposed [8]–[11]. Graphene is a material composed of a single layer of carbon atoms arranged in a hexagonal structure having exceptional properties such as optical transparency, flexibility, high electron mobility, and tunable conductivity, to name but a few. These features can be engineered using electrochemical potential energy [9], [12], [13]. Furthermore, graphene can support Surface Plasmon Polaritons (SPPs); thus, all in all, making it a prominent material for terahertz absorber applications [14]–[16].

Different two-dimensional graphene-based terahertz absorber structures have been proposed such as graphene stacks [17], square graphene patch [18], Periodic Arrays of Graphene ribbons (PAGRs) [19]–[21], and Periodic Arrays of Graphene Disks (PAGDs) [22], [23]. In broadband applications, a normalized absorption bandwidth of 100% has been reported for PAGRs configurations using three layers of graphene ribbons [21]. Although this number is far better than the other reports [17], [20], [24], the normalized absorption bandwidth of TM polarization has been enhanced to 131% using hybrid PAGRs and PAGDs [22]. Given that graphene disks absorb both TE and TM polarizations, the absorption of TE polarization has been enhanced. However, its normalized absorption bandwidth for both TM and TE is less than 70% which hinders its commercialization. Therefore, we need an ultra-broadband absorber with high normalized absorption bandwidth that absorbs both TE and TM polarization.

In this paper, a terahertz absorber based on three-layered graphene disk array has been proposed. Equivalent circuit model and TLM have been used for device design. Besides, the Finite Element Method (FEM) was also used to validate the findings. In the first section, the structure of the device has been demonstrated. In the following sections, the circuit model, design method, analysis, and discussion of the results has been presented.

## 2. Structure Design

The proposed structure depicted in Fig. 1 is composed of three layers of graphene disks with different radii which are separated with dielectric materials. The bottom layer of the structure is a sheet of gold (Au) with a short circuit functionality. The structure is designed to have an asymmetric Fabry-Perot-like configuration. The graphene disks are front mirrors, and the golden plate is the backside reflector. The incident wave in Terahertz regime is trapped and dissipated in the device. The dielectric material has chosen to be TOPAS polymer (Cyclic Olefin Copolymers) which has approximately 1.53 refractive index in terahertz band. This polymer can readily be fabricated and embedded in the structure via injection mold or extrusion method[25], [26]. Furthermore, its low loss in the THz spectrum makes it a promising candidate for broadband absorber application[26], [27]. In the current work, the graphene disks are assumed to have 1 nm thickness. As it is apparent from Fig. 1, d and p denote diameter and period of graphene disks, respectively, and the $h_s$ parameter is used for the thickness of different layers of TOPAS dielectric. The incident wave is assumed to have both TE and TM polarizations. Fig. 1. (b, c, d) show the top view of the first, second, and third layers, respectively.

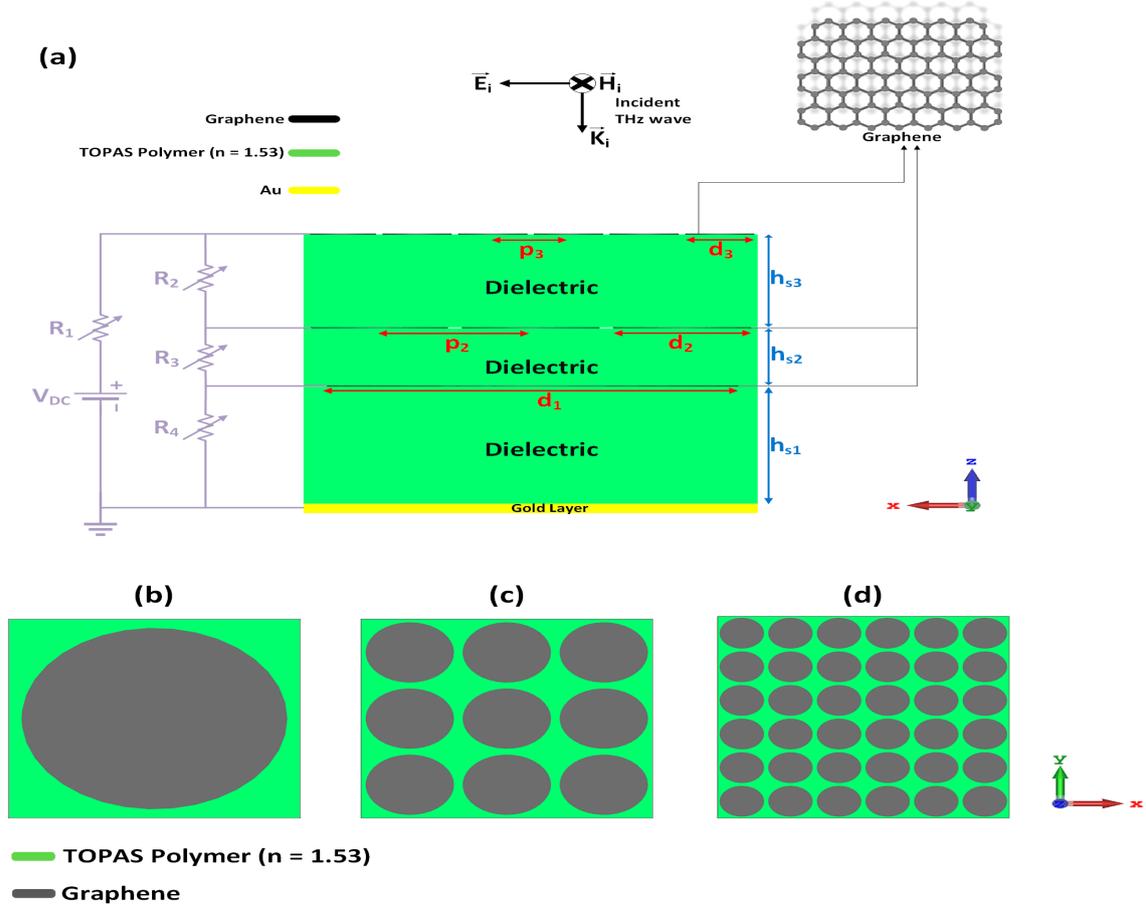

Fig. 1. The schematic structure of THz absorber based on the graphene disk array (a), the unit cell of first (b), second (c), and third (d) layers.

The conductivity of the graphene, which can be derived by Kubo formula, usually describes its properties [28]. This formula defines the conductivity of the graphene in a broad range of frequencies including Terahertz [18]. In the following, $\sigma_g$ consists of two parts: the intraband and interband contributions.

$$\sigma_g = \sigma_{\text{intra}} + \sigma_{\text{inter}} \tag{1}$$

$$\sigma_g = \frac{e^2 K_B T_K}{\pi \hbar^2 (\tau_g^{-1} + j\omega_g)} 2Ln[2\cosh(\mu_c / 2K_B T_K)] + \frac{e^2}{j4\pi\hbar} Ln[\frac{2\mu_c - \hbar(\omega - j\tau_g^{-1})}{2\mu_c + \hbar(\omega - j\tau_g^{-1})}] \tag{2}$$

Where $e$, $K_B$, $T_K$, $\hbar$, $\tau_g$, $\omega_g$, and $\mu_c$ are respectively electron charge, Boltzmann constant, temperature (fixed to 300 K), reduced plank constant, electron-phonon relaxation time, angular frequency, and chemical potential of graphene. From equation (2), it is obvious that if $\hbar\omega \ll \mu_c$, which is satisfied in THz frequency, the interband part can be neglected. Thus, we have:

$$\sigma_g = \frac{e^2 K_B T_K}{\pi \hbar^2 (\tau_g^{-1} + j\omega_g)} 2Ln[2\cosh(\mu_c / 2K_B T_K)] \tag{3}$$

It can be inferred that the conductivity (also the permittivity) of the graphene can be engineered with tuning the chemical potentials; i. e., we can change the absorption wavelength of the graphene by changing its chemical potential (Fermi level). The Graphene sheet can hold Surface Plasmon Polaritons, but its plasmon frequency is beyond the THz region. Thus, it is not suitable for THz absorber applications [29]. One of the prevalent mechanisms for tuning the plasmon frequency of the graphene sheet is making a pattern (grating) on it [30]–[32]. Making patterns on graphene sheet not only generates extra wave vectors (e.g. $\beta_n = 2n\pi/L$, where $K_n$ is $n$th wave vector, and $L$ is the grating period), but also it decreases the number of plasmons [31]. To have absorption, the incident wave must have a field component in the direction of the grating to make the electrons resonance. In this paper, we have used two-dimensional disk gratings. This configuration can hold both TE and TM polarizations due to its grating in both directions. The incident wave is assumed to have both TE and TM polarization with a normal incident angle. Therefore, we can expect that SPPs in both grating directions lead to absorption in intended frequencies.

The equivalent transmission line model of the device concerning TLM is shown in Fig. 2(a). Every element of this device can be modeled as an equivalent circuit component having its specific impedance. The metallic gold layer can be approximately modeled as a short circuit. The dielectric material can be considered as an element which has inductance or capacitance according to its thickness [33]. It has been shown that graphene disk arrays have resonance frequencies able to be modeled as RLC circuit components. Graphene disk arrays have eigenmodes that can be excited by specific incident frequencies. Also, it has been demonstrated that these eigenmodes are infinite and each can be modeled as a series RLC circuit parallel to the others [34]. Here, $Z_0$ is the free space impedance. $Z_{gi, i=1:3}$, the equivalent impedance of the RLC circuit (Fig. 2(b)), represents the impedance of each graphene layer. The impedance of dielectric material is shown by $Z_{si, i=1:3}$. We define $\beta_s$ and $Z_s$ as electromagnetic wave's propagation constant in dielectric intermediate material and impedance of dielectric material, respectively. The equivalent impedance for each part can be derived as follows:

$$Z_1 = jZ_{s1} \tan \beta_{s1} h_{s1} \tag{4}$$

$$Z_{i1} = \frac{Z_{g1} \cdot Z_1}{Z_{g1} + Z_1} \tag{5}$$

$$Z_2 = Z_{s2} \frac{Z_{i1} + jZ_{s2} \cdot \tan(\beta_{s2} h_{s2})}{Z_{s2} + jZ_{i1} \cdot \tan(\beta_{s2} h_{s2})} \tag{6}$$

$$Z_{i2} = \frac{Z_{g2} \cdot Z_2}{Z_{g2} + Z_2} \tag{7}$$

$$Z_3 = Z_{s3} \frac{Z_{i2} + jZ_{s3} \cdot \tan(\beta_{s3} h_{s3})}{Z_{s3} + jZ_{i2} \cdot \tan(\beta_{s3} h_{s3})} \tag{8}$$

$$Z_{in} = \frac{Z_{g3} \cdot Z_3}{Z_{g3} + Z_3} \tag{9}$$

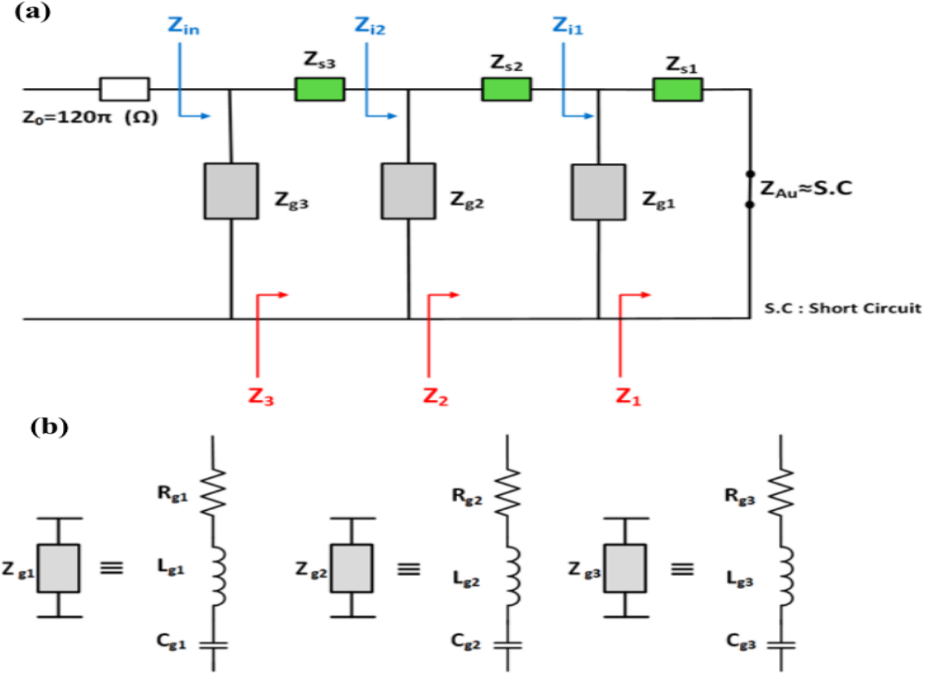

**Fig. 2. The transmission line model and impedance of each layer (a), the circuit model of graphene disks (b).**

Without losing generality, we consider only the first eigenmode for simplicity. Generally, the resistance, inductance, and capacitance for $n$th eigenmode are derived as follows [34]:

$$R_{g_n} = \frac{K_n p^2 \hbar^2}{S_n^2 e^2 \pi \mu_c \tau_g} \tag{10a}$$

$$L_{g_n} = \frac{K_n p^2 \hbar^2}{S_n^2 e^2 \pi \mu_c} \tag{10b}$$

$$C_{g_n} = \frac{\pi^2 S_n^2 \varepsilon_{eff}}{p^2 K_n q_{11}} \tag{10c}$$

$$\varepsilon_{eff} = \varepsilon_0 \frac{(n_1^2 + n_2^2)}{2} \tag{11}$$

where $S_n$ is the eigenfunction integral for the surface current equation of the graphene disk arrays which is given in Table 1 [34]. Also, $n_1$ and $n_2$ are the refractive indices of the surrounding material of graphene disks, where $q_{11}$ is the first eigenvalue for the equation of the surface current on the disks (values of which are given in Table 1 for different 2r/L). We have designed the PAGDs in the vicinity of its first resonance frequency. Therefore, for simplicity and without loss of generality, we have taken account only the first mode with $q_{11} \approx 0.4/r$, and d = 2r, and also $S_1$, $K_1$, and $\xi_{11}$ can be calculated as follow [34]:

$$S_1 = \int_0^r [\frac{df_{11}(\rho)}{d\rho}\rho + f_{11}(\rho)]d\rho \cong 0.6r , \quad K_1 = \int_s \xi_{11}.\xi_{11}^* dS \cong 1.3, \text{ and } \xi_{11} = \nabla[f_{11}(\rho)\cos\varphi] \tag{12}$$

**Table 1. Eigenvalues of the equation governing the surface current on graphene disks [34].**

| $d/p$ | 0.1 | 0.5 | 0.9 |
|---|---|---|---|
| $q_{11}d/2$ | 0.539 | 0.527 | 0.417 |

To have resonance in the circuit model and consequently a broadband absorber, the imaginary part of the equivalent impedance should be designed such that it gains a near zero number. Also, in order to have impedance match, the input impedance ($Z_{in}$) of the circuit should be approximately equal to the that of the free space ($Z_0 = 120\pi$). The resonance frequency of each graphene disk is derived from $f_0 = 1/2\pi\sqrt{LC}$. Thus, the graphene disk arrays would have only the real part. The other impedance effect stems from the dielectric materials. The thickness of the dielectric materials should be chosen such that the real part of the equivalent input impedance be approximately matched to that of free space.

## 3. Results and Discussion

Designing the TE and TM polarization-independent ultra-broadband absorber with the conditions mentioned in the previous section led to the structure for which the parameters are summarized in Table 2. The proposed structure has 1.78 THz central frequency with a bandwidth of 2.64 THz. The normalized bandwidth of the device is 148% which is to best of our knowledge the highest normalized absorption bandwidth among ever-reported graphene absorbers in the THz regime.

**Table 2. Summarized parameters of graphene disk arrays.**

| Layer | One | Two | Three |
|---|---|---|---|
| Graphene Disk Radius | $r_1 = 39.02$ μm | $r_2 = 12.95$ μm | $r_3 = 6.50$ μm |
| Graphene Disk Periods | $p_1 = 85.37$ μm | $p_2 = 28.46$ μm | $P_3 = 14.23$ μm |
| Height of Dielectric Layer | $h_{s1} = 24.56$ μm | $h_{s2} = 12.28$ μm | $h_{s3} = 19.65$ μm |
| Chemical Potentials | $\mu_{c1} = 0.96$ eV | $\mu_{c2} = 0.44$ eV | $\mu_{c3} = 0.18$ eV |

We have normalized the input impedance of the device to the free space impedance, and its real and imaginary part is depicted in the Fig. 3. It is apparent that in the central frequency and its proximity, not only the imaginary part is almost zero, but also it is flat without any sharp change. Also at these frequencies, the real part of the normalized input impedance is near unity which guarantees input impedance matching condition. These outstanding characteristics of proposed absorber result in an ultra-wideband THz absorber.

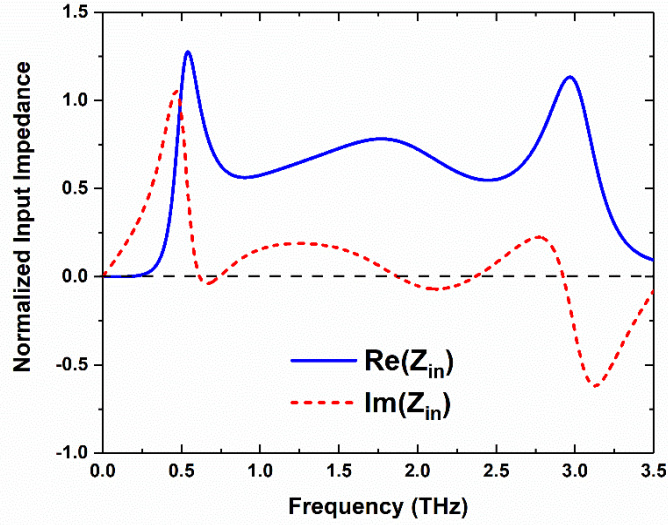

**Fig. 3. Normalized Input Impedance by TLM.**

To further analyze our proposed structure, we have performed full wave simulation using Finite Element Method (FEM) to validate our TLM. As it is obvious from Fig. 4, the two methods are in good agreement with each other. The bandwidth of the device is 2.64 THz ranging from 0.46 to 3.1 THz. The inset of the Fig. 4(a) shows the absorption over 90% of the device. Since the gold layer reflects the incident wave, the absorption can be determined by $A = 1 - |\Gamma|^2$, where $\Gamma$ is the return loss coefficient or reflection coefficient. Given that the absorption over 90% is a limit to consider a device to be an absorber, the return loss coefficient should be less than 32%. This parameter is shown in Fig. 4(b).

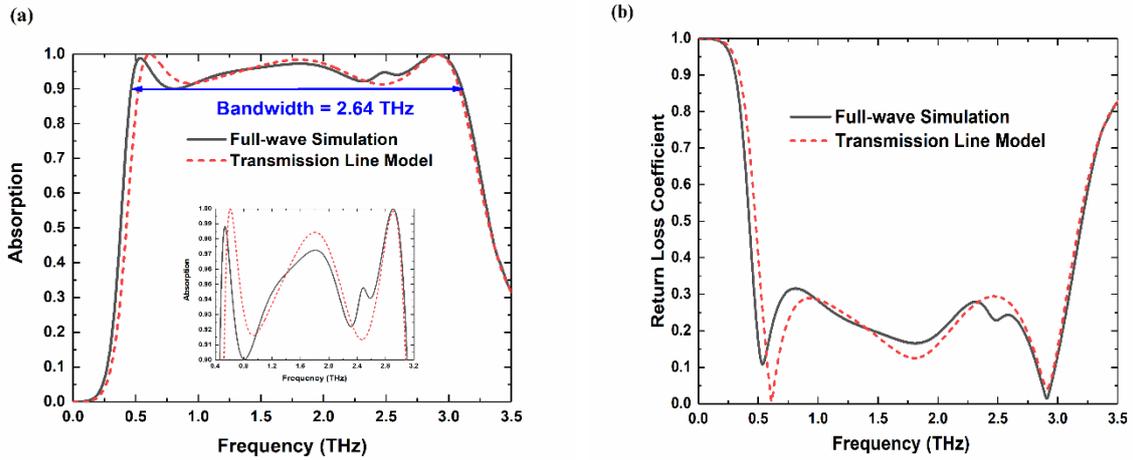

**Fig. 4 Absorption bandwidth (a), return loss coefficient (b) obtained with full-wave simulation and transmission line theory. The inset shows absorption over 90%.**

In this stage, we investigate the contribution of each layer for the total absorption, as in Fig. 5. The black curve is responsible for absorption of the device with only the first layer. It can be seen that absorption in the device is mostly due to this layer; however, it is below than 90% absorption indicating the necessity of other layers. The blue line showing absorption of the first and second layers exhibits some enhancement in the absorption, but the bandwidth is still at mediocre levels. We intended to design an ultra-broadband absorber with the highest normalized bandwidth absorption. Hence, we have added the third layer to realize this ambitious objective. The red line confirms our success. Besides, it is notable that graphene disks absorb not only TE but also the TM polarizations which is a great advantage. We can also infer that adding layers with smaller disks augments absorption at lower frequencies.

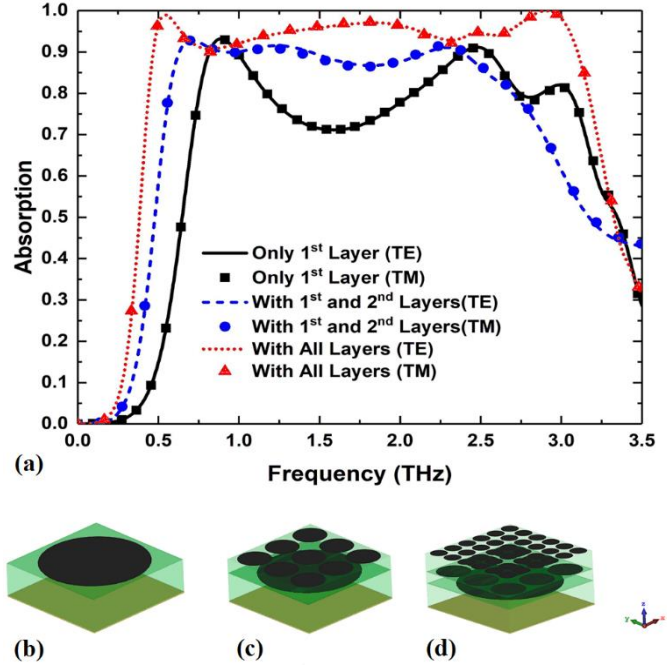

**Fig. 5. The absorption of only the first layer, the first two layers, all three layers (a), schematic view of only first layer (b), the first two layers (c), all three layers(d).**

Now, we study the effect of chemical potential contribution to absorption by applying it to one layer while making the other two layers have near zero chemical potential. The result is represented in Fig. 6(a). It is obvious the first layer is responsible for most of device's absorption except for the mid-range which is compensated by adding the other two layers. The third layer has absorption at both high and low frequencies, and it is added to increase overall device absorption to higher than 90% at the desired frequencies. We can infer that chemical potential applied to graphene disks can significantly alter the absorption. This phenomenon has been attributed to graphene's surface conductivity. When the chemical potential changes, the current distribution that graphene disks can hold changes resulting in variation of circuit components values. Therefore, the equivalent circuit impedance is no longer matched to that of free space [34].

Next, we examine the effect of altering chemical potential for each layer while keeping that of the other layers fixed. The results are represented in Fig. 6(b-d). Again, the impact of chemical potential in abortion can be vividly seen. For example, in Fig. 6(c) the chemical potential of the first and third layers are respectively fixed to 0.96 eV and 0.18 eV while changing the chemical potential of the first layer. With 0.44 eV chemical potential, the maximum absorption was reached. However, when we increase or decrease that value, device bandwidth degrades significantly. Hence, our device is optimally engineered to have maximum broadband absorption. Furthermore, we can note that our device can absorb both TE and TM polarizations which is one of its interesting features.

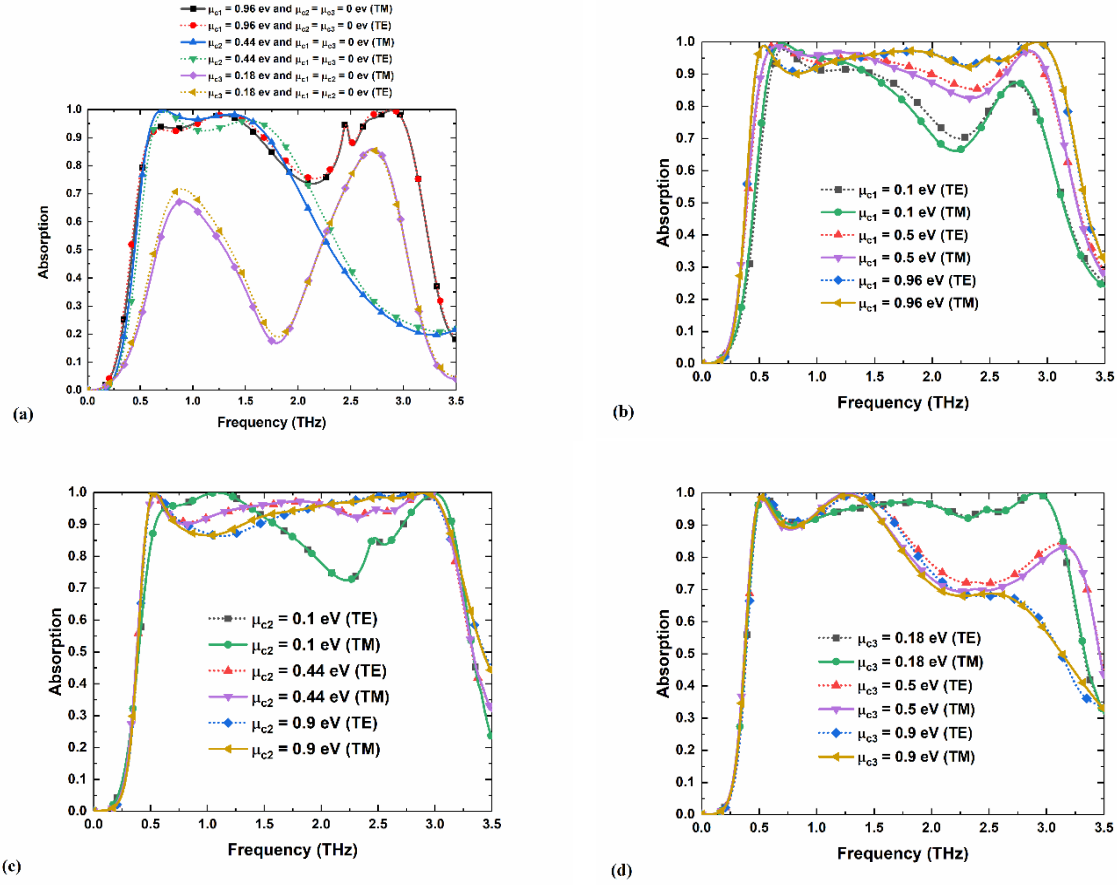

Fig. 6. The influence of applying chemical potential to only one layer (a), the influence of changing chemical energy of each layer while keeping others constant for TE and TM polarizations for the first (b), second (c), third(d) layer.

Dielectric thickness has an important effect on overall device performance. We have mentioned that the device functions as a Fabry-Perot resonator. The bottom gold layer reflects the incident wave and each graphene disk reflects parts of the frequency range; so, terahertz incident wave is trapped and dissipated in the device. Fig. 7(a-c) shows the effects of dielectric material's thickness on absorption. A significant variation in absorption can be seen by Fig. 7(a) varying the dielectric thickness. Changing $h_{s1}$ from 19.56 µm to 29.56 µm, the central frequency undergoes a redshift. Although the broadest bandwidth seems to be 19.56 µm, the absorption drops below 90% in 2 to 3 THz range. Thus, 24.96 µm becomes the best choice. For the other two layers, the redshift can be seen again with increasing the thickness. Nevertheless, its variation is not as notable as the first one. Considering TLM for the absorber, equations (4-8) indicate that the input impedance of the absorber can be tuned by dielectric thickness. From the input matching concept, the thickness of the first layer can be roughly derived from $h_{s1} = c_0 / 4 n_{s1} f_0$. By sweeping the thickness of the other two layers (matching overall input impedance to $120\pi$), the broadest absorption bandwidth has been obtained. From equation 2, it is obvious that the conductivity of graphene can change with temperature. Fig. 7(d) suggests that although the best performance of the device is in 300 ˚K, increasing or decreasing the temperature by 50 ˚K does not severely affects the absorption. This feature can be attributed to graphene's excellent thermal stability [35]–[37].

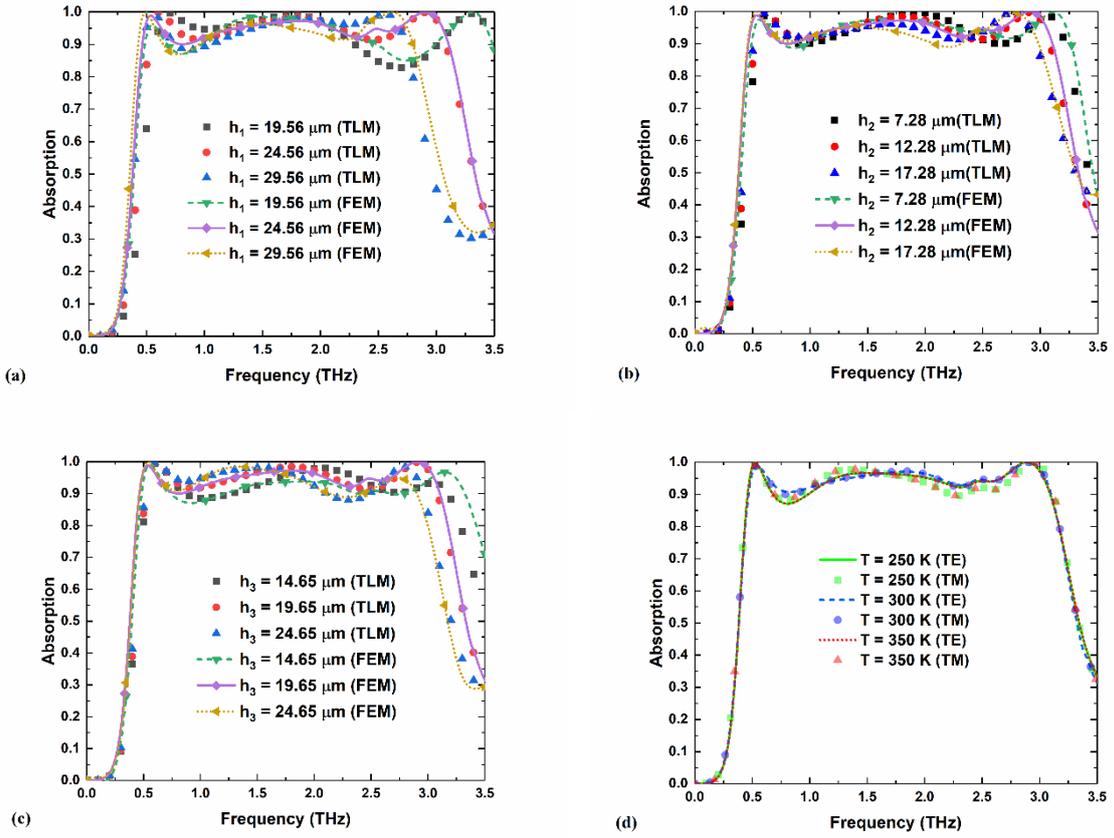

**Fig. 7. Effects of dielectric thickness variations for first (a), second (b), and third (c) layer calculated by TLM and FEM. Influence of thermal fluctuation of the proposed device (d).**

In the last attempt to analyze the device we investigate the voltage applied to the graphene disks. The term "ON" is used when the voltages are applied to the graphene disks and the term "OFF" is used for no voltage applied. Fig 8(a,b) show a remarkable change in the absorption. The absorption enhancement by applying voltage is due to the changes in surface conductivity of the graphene. The device independence to TE and TM polarizations is again apparent from Fig. 8.

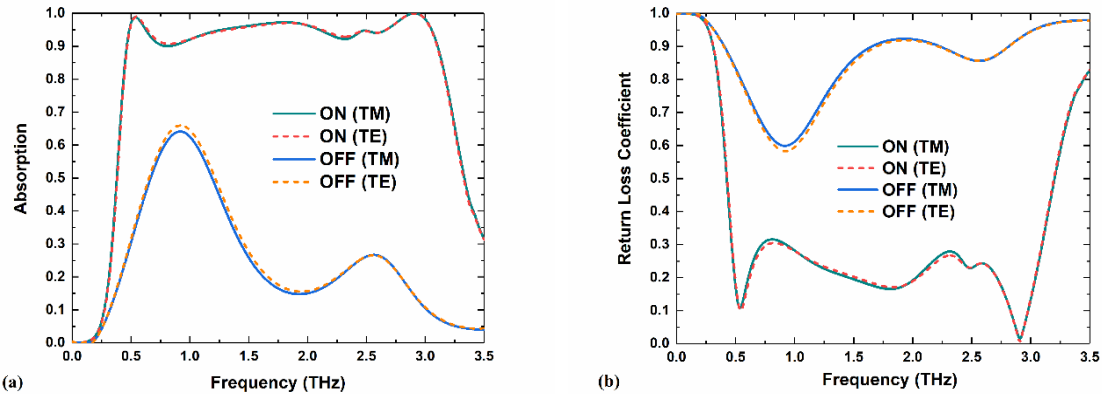

**Fig. 8. Absorption (a) and return loss coefficient (b) of the device under "ON" and "OFF" working conditions for both TE and TM polarizations.**

Fig. 9 shows electric field distribution in each layer of the device, which is obtained under the normalized incident wave at the central frequency for both TE and TM polarizations. We have applied positive potential energy to the graphene disk and a negative one (relative to potential applied to graphene disks). The gold layer functions as a backside reflector. Therefore, we can observe a 180 phase difference between gold and first graphene disk layer. Also, from figure 9(a, b) it is apparent that most of the electric field distribution and consequently the absorption is due to the first graphene disk layer as discussed above, and the absorption in the second and third layers diminishes gradually.

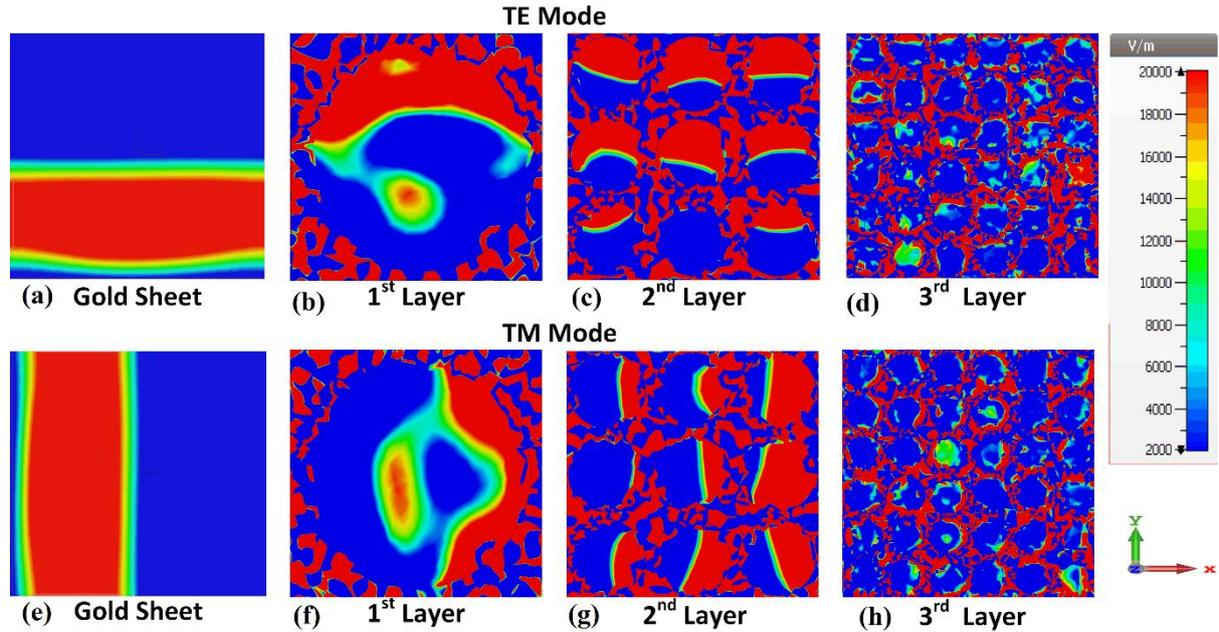

**Fig. 9. The electric field distribution in gold (a, e), the first (b, f), the second (c, g), and the third (d, h) layers for TE and TM modes, respectively**.

It is worth to overview and compare the reported results for ultra-broadband absorbers with the proposed structure. Our proposed structure reached 148% normalized bandwidth absorption ($BW/f_0$) which is the highest amount in graphene THz absorbers to the best of our knowledge. Another advantage of our structure is that it absorbs both TE and TM polarizations. Also, it is a layered structure, and it can be fabricated using common methods (e.g., Chemical Vapor Deposition).

**Table 3. Comparison of proposed structure with other graphene-based absorbers.**

| Ref | Number of Graphene layers | Device Height (µm) | $f_0$ (THz) | Polarization | $BW/f_0$ |
|---|---|---|---|---|---|
| [24] | 3 | 11 | 6.3 | TE and TM | 25.3% |
| [17] | 5 | 4.5 | 3.07 | TE and TM | 31.09% |
| [20] | 9 | 28.5 | 5.39 | TM only | 88% |
| [19] | 1 | 50 | 1 | TM only | 98% |
| [21] | 3 | 26.67 | 2.54 | TM only | 100% |
| [22] | 3 | 61 | 1.67 | TM only | 131% |
| **Present Work** | **3** | **56.49** | **1.78** | **TE and TM** | **148%** |

## 4. Conclusion

In this work, we have proposed ultra-broadband, high, and normalized bandwidth terahertz absorber with graphene disk arrays. The device design was performed using the analytical circuit model of the structure with the help of TLM. Maximum absorption was achieved with the tuning input impedance of the device to be matched to that of free space. Our design using transmission line theory was validated with the Finite Element Method (FEM). Also, we have shown the device's thermal stability for at least 100 ˚k temperature fluctuation in working conditions. Furthermore, the proposed structure absorbed both TE and TM polarizations due to graphene disks symmetry. All of the considerations made led to obtaining 148% normalized bandwidth absorption. To the best of our knowledge, the proposed structure has the highest normalized absorption bandwidth among two-dimensional graphene-based THz absorbers. Comparing with other metamaterial absorbers, there is only one report better than ours with 152% normalized absorption bandwidth [38]. Although it is a better result, the central frequency of the absorber is lower (1.42 THz) than our proposed one (1.78 THz). Given that designing absorber at low frequencies is much easier, the proposed structure may be marked as the best report.